\documentclass[twocolumn]{aastex631}

\usepackage{color}
\usepackage{amsmath}
\usepackage{xcolor}
\usepackage{mathrsfs}
\usepackage{natbib}
\usepackage{gensymb}
\usepackage{appendix}
\usepackage[T1]{fontenc}

\shortauthors{Bryan $\&$ Lee}
\shorttitle{GG/SE correlation}

\begin{document}

\title{Resolving the Super-Earth/Gas Giant Connection in Stellar Mass and Metallicity}

\author[0000-0002-6076-5967]{Marta L. Bryan}
\affiliation{David A. Dunlap Department of Astronomy $\&$ Astrophysics, University of Toronto, 50 St. George St., M5S 3H4 Toronto ON, Canada}
\affiliation{Department of Chemical $\&$ Physical Sciences, University of Toronto Mississauga, 3359 Mississauga Road, L5L 1C6 Mississauga ON, Canada}

\author[0000-0002-1228-9820]{Eve J.~Lee}
\affiliation{Department of Physics and Trottier Space Institute, McGill University, 3600 rue University, H3A 2T8 Montr\'eal QC, Canada}
\affiliation{Trottier Institute for Research on Exoplanets (iREx), Universit\'e de Montr\'eal, Canada}
\affiliation{Department of Astronomy \& Astrophysics, University of California, San Diego, La Jolla, CA 92093-0424, USA}

\begin{abstract}

The observed correlation between inner super-Earths and outer gas giants places strong constraints on formation theories. Building on previous work, Bryan $\&$ Lee 2024 showed that there is a statistically significant positive correlation between super-Earths and outer gas giants around metal-rich FGK stars, and that this correlation disappears for metal-poor hosts. Here we consider how this connection evolves across stellar mass. Starting with our sample of 85 M-dwarfs ($<$0.6 M$_{\odot}$) hosting inner super-Earths, we calculate P(GG|SE, [Fe/H]$>$0) = 9.4 (+10.2 -3.1)$\%$ and P(GG|SE, [Fe/H]$\leq$0)$<$3.1$\%$. Compared to the field gas giant frequency calculated from the Rosenthal et al 2021 sample, we find P(GG|[Fe/H]$>$0) = 10.3 (+6.9 -3.1)$\%$, and P(GG|[Fe/H]$\leq$0)$<$2.6$\%$ for M-dwarfs. While we see a higher gas giant frequency around metal-rich M-dwarfs for both samples, we find no significant correlations between super-Earths and gas giants. Combining our 85 M-dwarf sample with our FGK sample from Bryan $\&$ Lee 2024, we resolve the SE/GG correlation in stellar mass (0.3--1.5 M$_{\odot}$) and metallicity. We show the positive correlation emerges in metal-rich K-dwarfs and strengthens with increasing stellar mass. Gas giant properties also impact the correlation -- for metal rich stars, the positive correlation is strengthened by: 1) dynamically hot gas giants for all stellar masses; 2) distant gas giants only for higher mass stars; and 3) single gas giants for K-dwarfs and multiple gas giants around more massive stars. We discuss how the stellar mass dependence of the inner-outer planet correlation can be understood from the increasing disk mass budget for higher mass stars. 

\end{abstract}

\section{Introduction}

Given the dominant role Jupiter and Saturn likely played in the evolution of our Solar System, it is important to investigate the impact extrasolar gas giants have on the lives of inner extrasolar systems \citep{Morbidelli07,Walsh11,Batygin15}. Thus far the connection between gas giants and inner systems has primarily focused on systems hosting inner super-Earths, small planets 1--20 M$_{\earth}$ and 1--4 R$_{\earth}$. Early studies by \citet{ZhuWu18} and \citet{Bryan2019} found a tentative positive correlation between super-Earths and Jupiters, indicating Jupiters do not play an adversarial role in the formation of super-Earths and instead reflect favorable conditions for the emergence of both planet populations. While there was subsequent debate over this correlation, with some studies tentatively finding no or even negative correlations \citep[e.g.][]{Bonomo2023}, follow-up work demonstrated these diverging results arose from a combination of biased sample selection and small sample sizes. In particular, \citet{Zhu2023} found that when just taking the metal-rich [Fe/H]$>$0 systems from \citet{ZhuWu18}, \citet{Bryan2019}, and \citet{Bonomo2023}, all studies get consistent answers -- tentative positive correlations. 

Recently, in \citet{Bryan2024} (hereafter BL24) we reexamined this correlation and the emerging dependence on host star metallicity with a sample of 184 super-Earth systems, three times the size of the sample of \citet{Bryan2019}. We found a significant positive correlation between super-Earths and gas giants around metal-rich stars, and that this correlation is enhanced for systems where the gas giant is more distant ($>$3 AU), more eccentric ($e>$0.2), and/or in multi-gas giant systems. This positive correlation disappears for metal poor [Fe/H]$\leq$0 systems. Taken together, these findings corroborate the conclusion of \citet{Zhu2023} and support the critical role that metallicity, as a proxy for disk solid content, plays in shaping the planetary architectures. Namely, disk solid content establishes systems with outer giant and inner small planets \citep[e.g.,][]{Chachan22,Chachan23} given that most solid mass will drift to the inner region during the assembly of a Jupiter-nucleating cores \citep[e.g.,][]{Lin18} so that there is enough mass in the inner disk even after an outer giant completes its formation.

These previous studies have focused on systems around sun-like stars. Here we turn our focus to lower-mass stars, exploring the connection between super-Earths and gas giants around M stars ($<$0.6 M$_{\odot}$). In Section \ref{section: obs} we describe our sample of 85 M-dwarf systems hosting inner super-Earths (see Table \ref{tab:full-sample}), as well as our comparison sample for field star gas giant occurrence rates. Section 3 details how we calculate individual system sensitivities to distant gas giants, and Section 4 shows how those sensitivities are incorporated into the occurrence rate calculation. Section 5 explores the dependence of this correlation in mass and metallicity space by combining our sample of 85 M-dwarf super-Earth systems with our previous sample of 184 super-Earth systems around higher-mass host stars. Finally, section 6 places these findings in the context of planet formation theory.

\section{Observations}
\label{section: obs}

Similar to BL24, we build our M-dwarf sample using five criteria: the system has 1) publicly available radial velocity (RV) datasets, 2) at least one confirmed super-Earth planet (1--20 M$_{\earth}$, 1--4 R$_{\earth}$), 3) host star mass $\leq$0.6 M$_{\odot}$, 4) RV datasets with over one year time baseline and more than 20 data points, and 5) an RV semi-amplitude K$>$1 m/s in the case of systems with super-Earths discovered using RVs, limiting the inclusion of false positive planets. Note the key difference between the sample of systems in this paper versus BL24 is the host star mass cutoff; BL24 selected stars $>$0.6 M$_{\odot}$, here we initially consider the lower mass sample $\leq$0.6 M$_{\odot}$. 

This selection yields a total of 85 super-Earth systems around M-dwarfs (see Table \ref{tab:full-sample}). Super-Earths in this sample span orbital periods $\sim$1 -- 3000 days. Of these 85 M-dwarf systems, 24 are metal-rich [Fe/H]$>$0, and 61 are metal-poor [Fe/H]$\leq$0. Only two of these systems host gas giants (0.5--20 M$_{\rm Jup}$), GJ 876 and HIP 22627. Both systems are metal-rich, [Fe/H] = 0.21$\pm$0.06 and [Fe/H] = 0.30$\pm$0.10, with stellar masses 0.37$\pm$0.01 M$_{\odot}$ and 0.36$\pm$0.03 M$_{\odot}$ respectively. GJ 876 has two gas giants, masses 2.11$\pm$0.04 M$_{\rm Jup}$ and 0.70$\pm$0.01 M$_{\rm Jup}$, and semi-major axes 0.218$\pm$0.002 AU and 0.136$\pm$0.001 AU respectively.
\footnote{We note that in addition to this system containing a super-Earth orbiting inside two gas giants, there is also an additional super-Earth orbiting outside the warm Jupiters. Across our stellar samples FGKM, only GJ 876 and WASP-47 have gas giant(s) interior to super-Earth(s).} HIP 22627 has one gas giant with mass 0.75$\pm$0.04 M$_{\rm Jup}$ and semi-major axis 2.52$\pm$0.03 AU. 

\begin{figure*}
 %   \centering
    \begin{tabular}{cc}
    \textbf{R21} &  \textbf{This Letter}\\
    \includegraphics[width=0.5\textwidth]{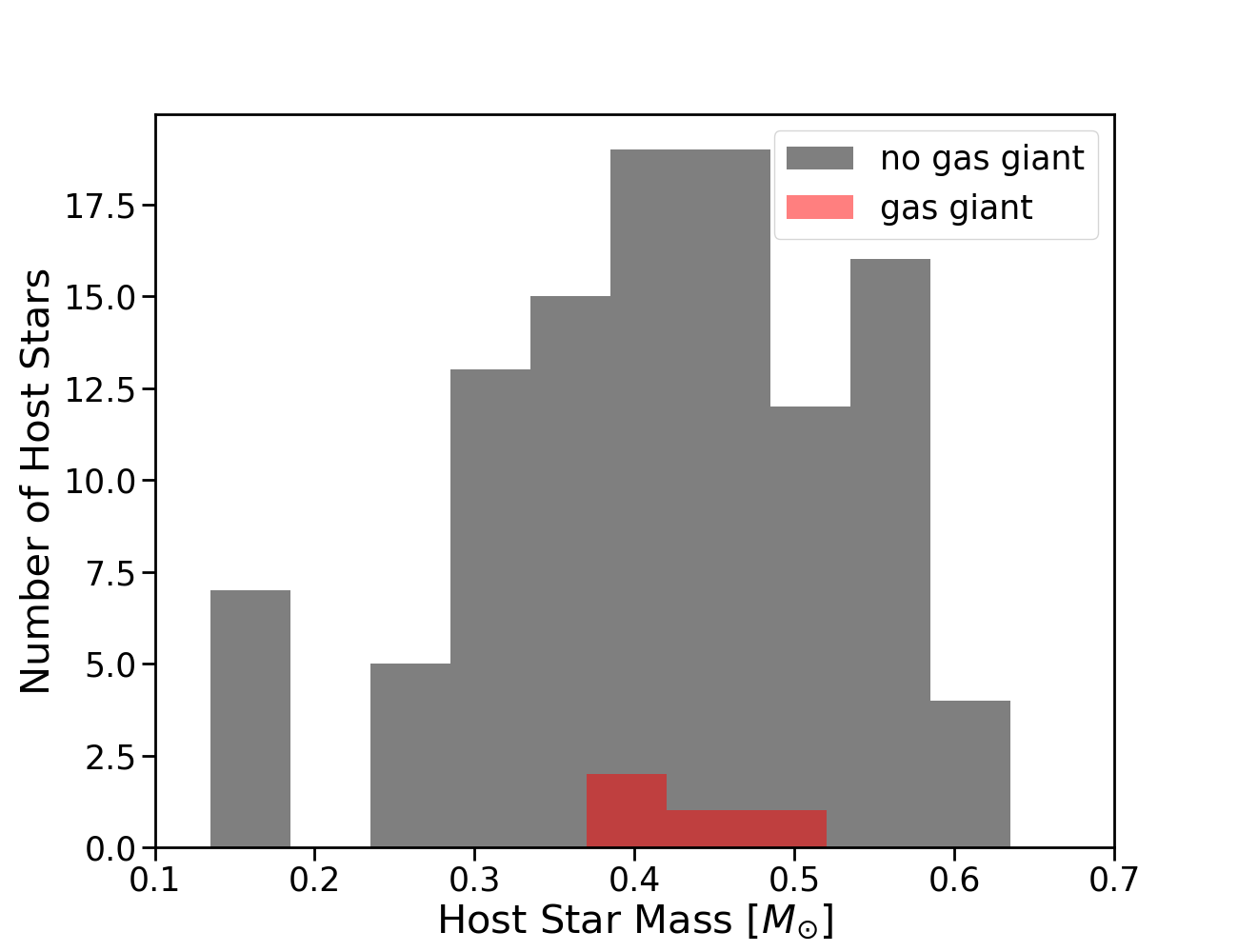} &

    \includegraphics[width=0.5\textwidth]{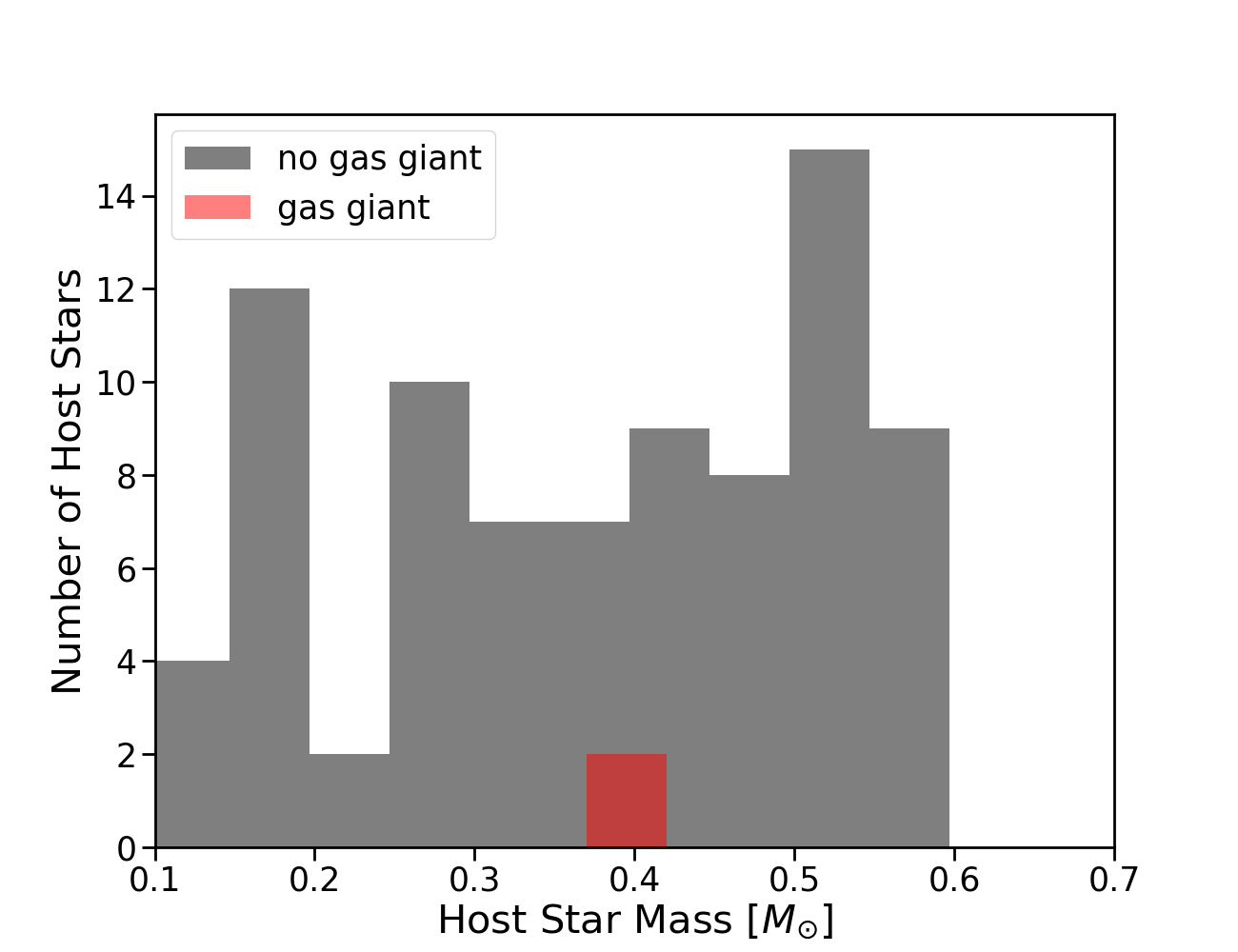}\\

    \includegraphics[width=0.5\textwidth]{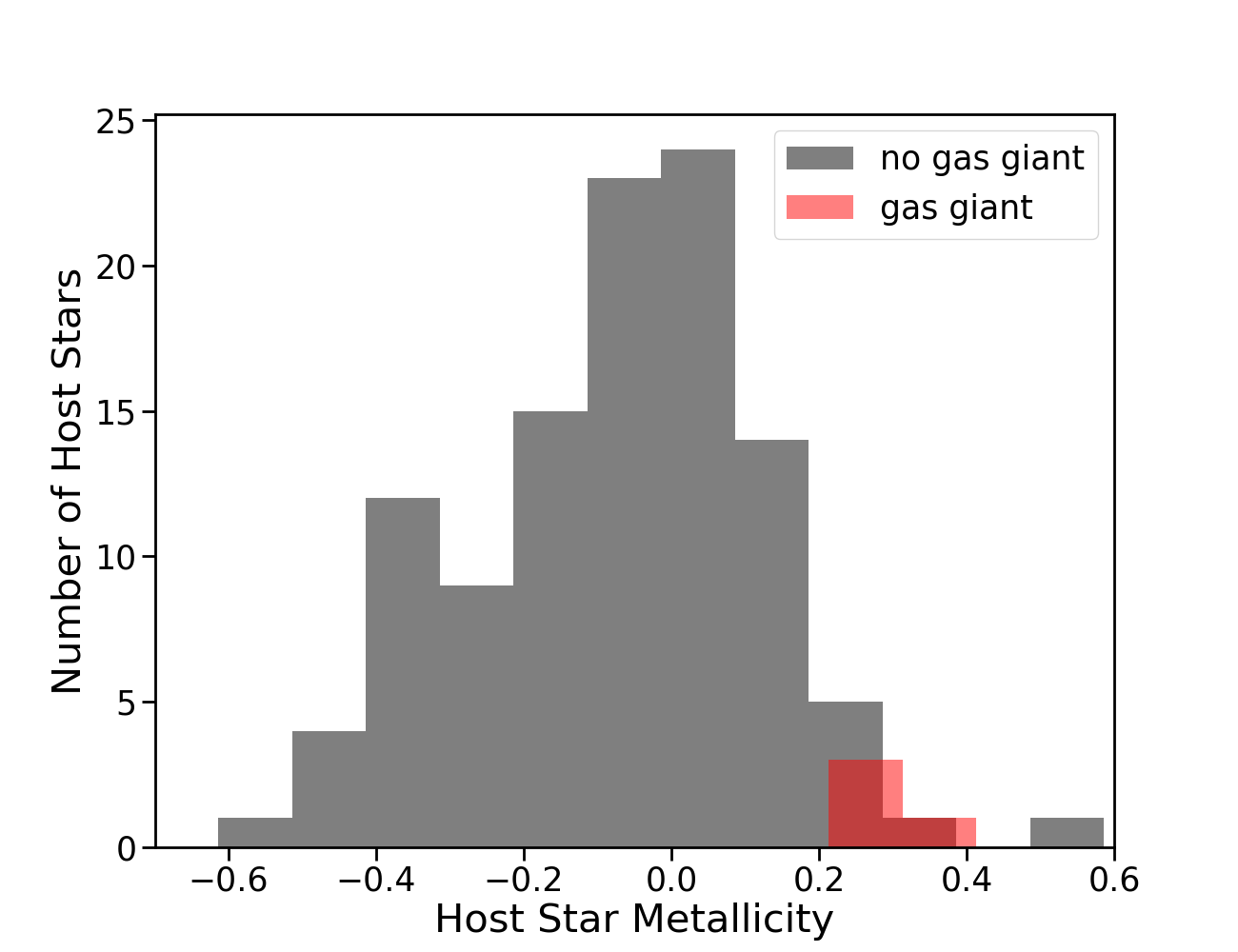} &
    \includegraphics[width=0.5\textwidth]{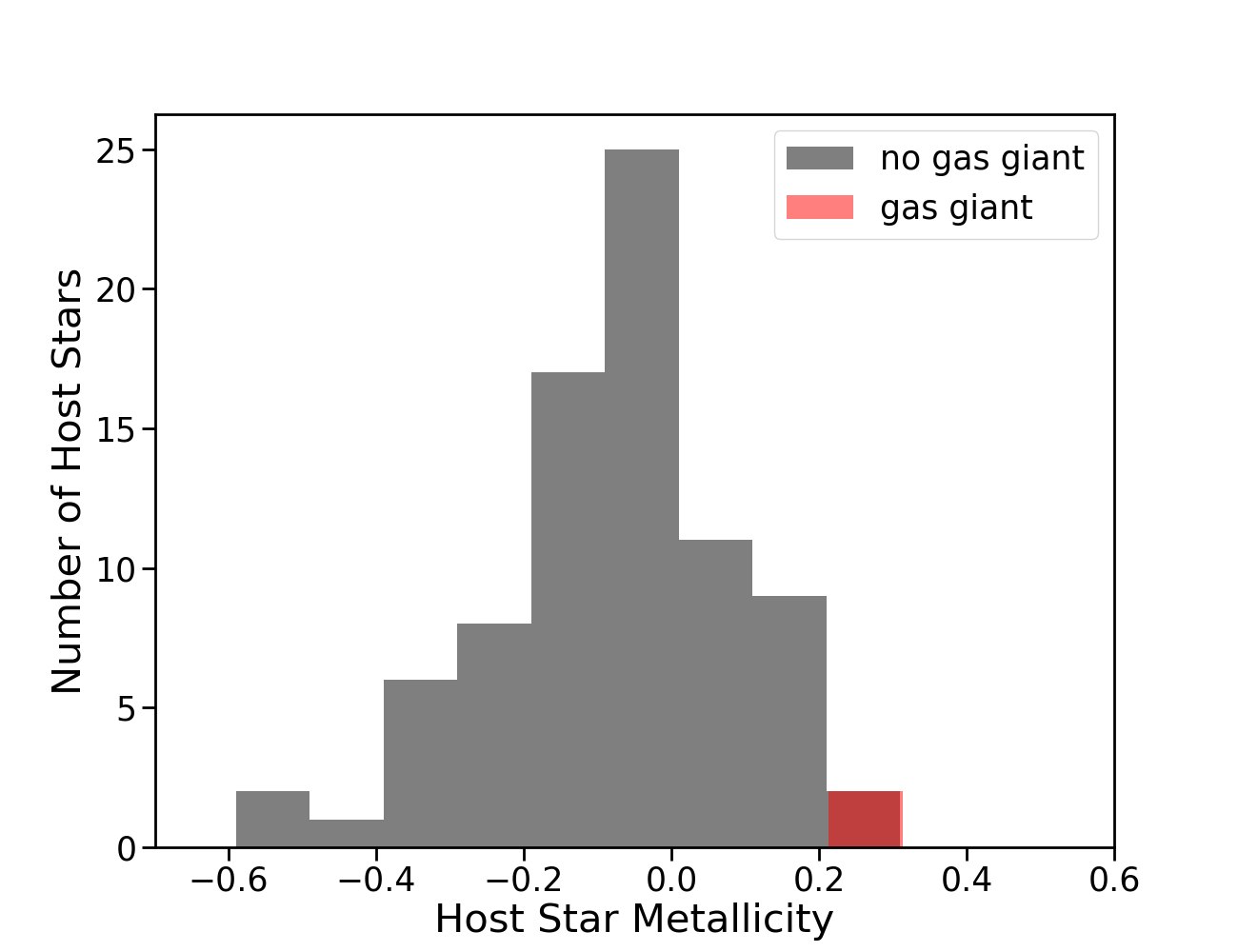}\\
    \end{tabular}
    \caption{Top left: M-dwarf mass distribution from the R21 sample. Top right: M-dwarf mass distributions from this paper. Bottom left: M-dwarf metallicity distribution from the R21 sample. Bottom right: M-dwarf metallicity distribution from this paper's sample. For all panels, black histograms correspond to systems without gas giants, red correspond to those with gas giants. Gas giants considered are 0.5$-$20 M$_{\rm Jup}$ and $<$ 10 AU. All systems in this paper's sample host inner super-Earths.}
    \label{fig:comp stellar dist}
\end{figure*}

To determine whether there is a positive, negative, or no correlation between super-Earths and gas giants in our sample, we need to compare the conditional occurrence rate P(GG|SE, M$_{\star}\leq 0.6 M_{\odot}$) that we can target with our sample, to the field occurrence of gas giants around M stars P(GG|M$_{\star}\leq 0.6 M_{\odot}$). For the field occurrence rate, we use systems from \citet{Rosenthal2021} (hereafter R21), which presents a catalog of RVs for 719 stars observed over the course of 30 years. Here we consider just the M stars from the R21 sample, totaling 114 systems with M$_{\star}\leq 0.6$ M$_{\odot}$. Of these, 41 are metal-rich [Fe/H]$>$0, and 73 are metal-poor [Fe/H]$\leq$0. Four systems have gas giants, GJ 317, HIP 109388, GJ 876, and HIP 22627, all of which are metal rich. Both GJ 317 and HIP 109388 have two gas giants each, masses 1.85$\pm$0.04/1.7$\pm$0.08 M$_{\rm Jup}$ and 0.89$\pm$0.04/1.08$\pm$0.05 M$_{\rm Jup}$, and semi-major axes 1.18$\pm$0.01/5.78$^{+0.74}_{-0.38}$ AU and 2.41$\pm$0.02/4.97$^{+0.08}_{-0.07}$ AU respectively. Properties of the gas giants in GJ 876 and HIP 22627 are listed in the previous paragraph, as they are included in our M-dwarf super-Earth sample. We show the distribution of mass and metallicity of the M-dwarf host stars for our super-Earth sample as well as the R21 sample in Figure \ref{fig:comp stellar dist}, highlighting how the small subset of M dwarfs hosting at least one outer gas giant has supersolar metallicity.

\section{Sensitivity Maps}
\label{sensitivity}
Like our sample of higher-mass host star systems in BL24, in this paper we have a heterogeneous sample of systems and datasets where individual system sensitivities to outer gas giants can vary significantly depending on time baseline, number of data points, observing cadence, and measurement precision. To quantify these differences in sensitivity, for each system we calculate completeness as a function of planet mass and semi-major axis given public RV datasets (see Table \ref{tab:full-sample} for references). Starting with a 50$\times$50 grid in mass and semi-major axis spaced evenly in log space from 0.3--30 M$_{\rm Jup}$ and 0.1--30 AU, we inject 50 simulated planets into every grid box. For each injected planet, we draw a mass and semi-major axis value from a uniform distribution across each grid box, an inclination $i$ from a uniform distribution in $\cos i$, an eccentricity from the $\beta$ distribution, and remaining orbital elements from uniform distributions. 

For each injected planet in a given system, at every observational epoch we calculate simulated RVs, adding noise by drawing from a Gaussian distribution with width defined by the measurement uncertainty randomly shuffled from the original host star RV dataset. To evaluate if an injected planet would be detected given the public dataset, we 1) fit each RV series with a one-planet orbital solution and a flat line; 2) calculate the Bayesian information criterion (BIC) values for each fit; 3) determine that a simulated planet is `detected' if the BIC value for the one-planet model fit is smaller than the BIC value for the flat line fit by more than 10 \citep{Jeffreys1939}. Alternatively, if either the flat line is preferred or the one planet model is preferred by a $\Delta$BIC$<$10, we consider the injected planet `not detected'. We repeat steps 1-3 for each injected planet, and use the `detected' and `non-detected' results to calculate completeness across the full mass/semi-major axis grid.\footnote{We note that the parameter space where sensitivity begins to decrease typically coincides with the time baseline spanning less than a full orbital period of the injected planet. Once there are only partial orbits, it is challenging to determine whether orbital motion is caused by just one planet or by a combination of several. In this case, model selection prefers the less complicated one-planet model. While we could include more than one injected gas giant, given our model selection criteria for a detection versus non-detection, we do not anticipate the sensitivity maps changing significantly.}

We compute sensitivity maps for each system in our M-dwarf super-Earth sample, as well as the M-dwarf R21 sample. Figure \ref{fig:ave completeness} illustrates the average completeness maps for our sample (right) and the R21 sample (left). Gas giants detected in each sample are overplotted. While the R21 sample on average has higher sensitivities to distant gas giants, we see that for both samples the detected gas giants fall in a mass/semi-major axis parameter space that has high if not 100$\%$ completeness. These sensitivity maps motivate our chosen mass and semi-major axis ranges of 0.1--10AU and 0.5--20 M$_{\rm Jup}$ in our gas giant occurrence rate calculations presented in the following section. We also note that while different observational biases can lead to differences in sensitivity for the super-Earth systems detected via transits versus RVs, as in BL24 we find consistent occurrence rates between the two sets of systems, and continue with the combined M-dwarf super-Earth sample below.

\section{No Correlation Between Super-Earths and Gas Giants Around M-Dwarfs}
\label{occ rate}

To explore the connection between super-Earths and gas giants around M-dwarfs, we want to calculate 1) the frequency of gas giants in super-Earth systems P(GG|SE) using the M-dwarf sample presented here; and 2) the frequency of gas giants around field M-dwarfs P(GG) from the R21 sample. Given the dependence of this SE/GG connection on metallicity previously demonstrated in BL24 and \citet{Zhu2023}, we split these M-dwarf samples into metal-rich [Fe/H]$>$0 and metal-poor [Fe/H]$\leq$0 systems. We subsequently calculate the frequency of gas giants (0.5-20 M$_{\rm Jup}$, 0.1-10 AU) in super-Earth systems around M-dwarfs P(GG|SE, [Fe/H]$>$0, M$_{\star}\leq 0.6$ M$_{\odot}$), and this same frequency around metal-poor M-dwarfs P(GG|SE, [Fe/H]$\leq$0, M$_{\star}\leq0.6$M$_{\odot}$), using a general binomial distribution:

\begin{equation}
    f(x;a,b) = \frac{1}{B(a,b)}x^{a-1}(1-x)^{b-1}
\end{equation}

\noindent where $B$ is the beta function, $a=n_{\rm det}+1$, $b=(n_{\rm eff}-n_{\rm det})+1$, $n_{\rm det}$ is the number of detections, e.g. systems with gas giants 0.5-20 M$_{\rm Jup}$ and 0.1-10 AU, and $n_{\rm eff}$ is the total number of systems modified by the sample completeness. To calculate $n_{\rm eff}$ and correct for individual system sensitivities, we took the completeness maps calculated for each system, summed over the 0.5--20 M$_{\rm Jup}$ and 0.1--10 AU integration range, and determined the average completeness for the set of systems considered. We find that metal-rich M-dwarfs have P(GG|SE, [Fe/H]$>$0, M$_{\star}\leq 0.6$ M$_{\odot}$) = 9.4 (+10.2 -3.1)$\%$, and with no detections metal-poor M-dwarfs have a 1$\sigma$ upper limit of P(GG|SE, [Fe/H]$\leq$0, M$_{\star}\leq0.6$M$_{\odot}$) $<$ 3.1$\%$.

\begin{figure}
\centering
\includegraphics[width=0.5\textwidth]{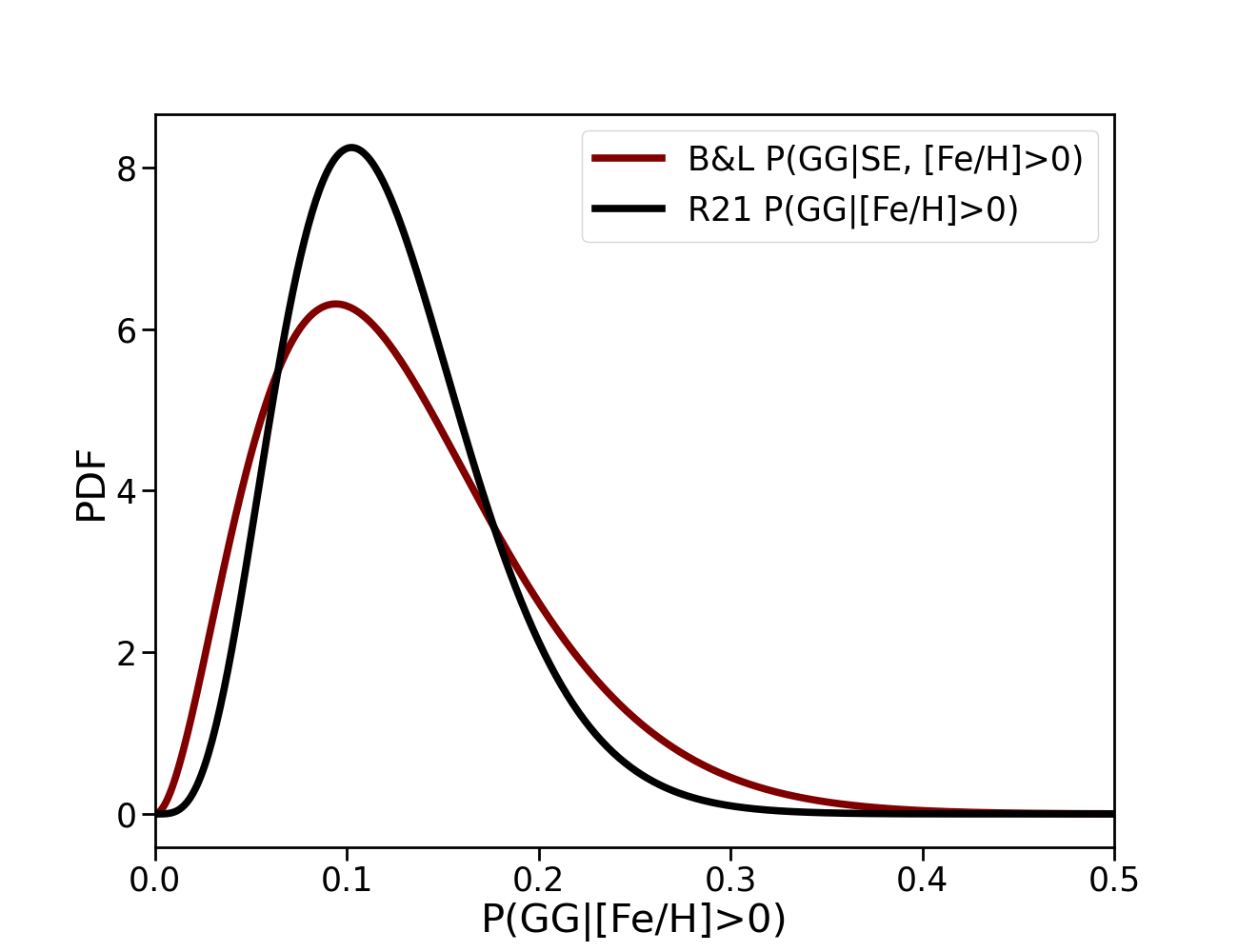}
\includegraphics[width=0.5\textwidth]{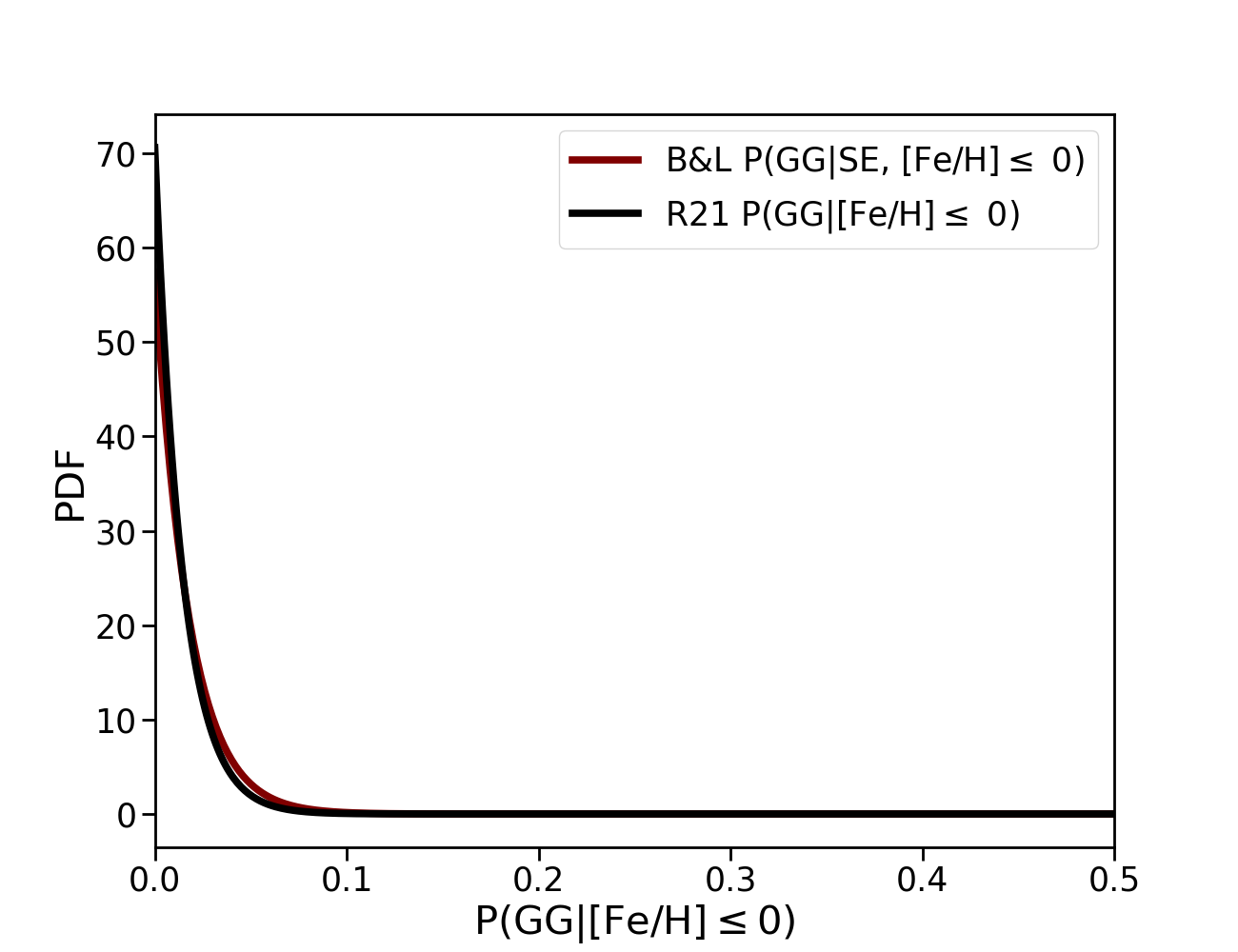}
\caption{Top: occurrence rate of gas giants in metal-rich M-dwarf systems. Bottom: same in metal-poor systems. Frequencies from this paper are in maroon, R21 are in black. While metal-rich M-dwarfs have a higher frequency of gas giants for both samples, there is no significant difference in occurrence rates between the samples. We find no significant correlation between super-Earths and gas giants around M-dwarf host stars.}
\label{fig: occ rate}
\end{figure}

We calculate P(GG) from the R21 sample using the same sensitivity calculation methods described in section \ref{sensitivity} and the same definition of a gas giant. This yields P(GG|[Fe/H]$>$0, M$_{\star}\leq 0.6$ M$_{\odot}$) = 10.3$^{+6.9}_{-3.1}\%$ and a 1$\sigma$ upper limit for metal poor M-dwarfs of P(GG|SE, [Fe/H]$\leq$0, M$_{\star}\leq 0.6$ M$_{\odot}$) $<$ 2.6$\%$. Figure \ref{fig: occ rate} illustrates these results. While we see a higher gas giant frequency around metal-rich M-dwarfs for both samples, we find no significant difference in gas giant occurrence rate between P(GG) and P(GG|SE). There is no significant correlation, positive or negative, between super-Earths and gas giants around M-dwarf host stars.

To further resolve the dependence on metallicity, as in BL24 we break the two samples of systems into three metallicity bins: low metallicity [Fe/H]$<$-0.1, mid metallicity -0.1$\leq$[Fe/H]$\leq$0.1, and high metallicity [Fe/H]$>$0.1. For both our sample and the R21 sample, all gas giant detections are in the high metallicity bin, corresponding to frequencies of P(GG|SE, [Fe/H]$>$0.1) = 18.3$^{+16.4}_{-6.4}\%$ and P(GG|[Fe/H]$>$0.1) = 20.0$^{+11.5}_{-6.0}\%$. While both frequencies are larger in this high metallicity bin, they remain consistent with each other.

\section{A Mass/Metallicity Transition to a Positive SE/GG Correlation}
\label{compare mstar sun}

While we find no significant correlation between super-Earths and outer gas giants around M-dwarfs, previous work has found a significant positive correlation between these planet populations around more massive stars that are metal rich (BL24). In this section we further explore how this correlation evolves in stellar mass and metallicity space.

We combine our super-Earth samples from BL24 and this paper, and divide this expanded set of systems into four mass bins: 0.3--0.55 M$_{\odot}$, 0.55--0.8 M$_{\odot}$, 0.8--1.05 M$_{\odot}$, and $>$1.5 M$_{\odot}$, broadly corresponding to early M, K, G, and F spectral types. To resolve the correlation in mass and metallicity space, we divide each mass bin into three metallicity bins: [Fe/H]$<$-0.1, -0.1$\leq$[Fe/H]$\leq$0.1, and [Fe/H]$>$0.1. We do the same to the R21 sample. Note that we leave out systems with host star masses less than 0.3 M$_{\odot}$ since the R21 sample does not have any systems in that mass range with metallicities [Fe/H]$>$0.1. 

We additionally include the sample of systems from \citet{Wittenmyer2020} in this analysis (hereafter W20). This sample consists of 203 predominantly FGK stars monitored over an 18-year period by the Anglo-Australian Planet Search. While the W20 sample has fewer systems than R21, we opt to include this additional P(GG) source for completeness and direct comparison with previous studies which have relied on W20 as a comparison sample to test the SE/GG correlation. We divide the W20 sample into the same mass and metallicity bins as above. Since the W20 sample does not include metal-rich M-dwarfs, we only use this sample for the three higher mass bins.

\begin{figure*}
%\centering
    \begin{tabular}{cc}
    \includegraphics[width=0.5\textwidth]{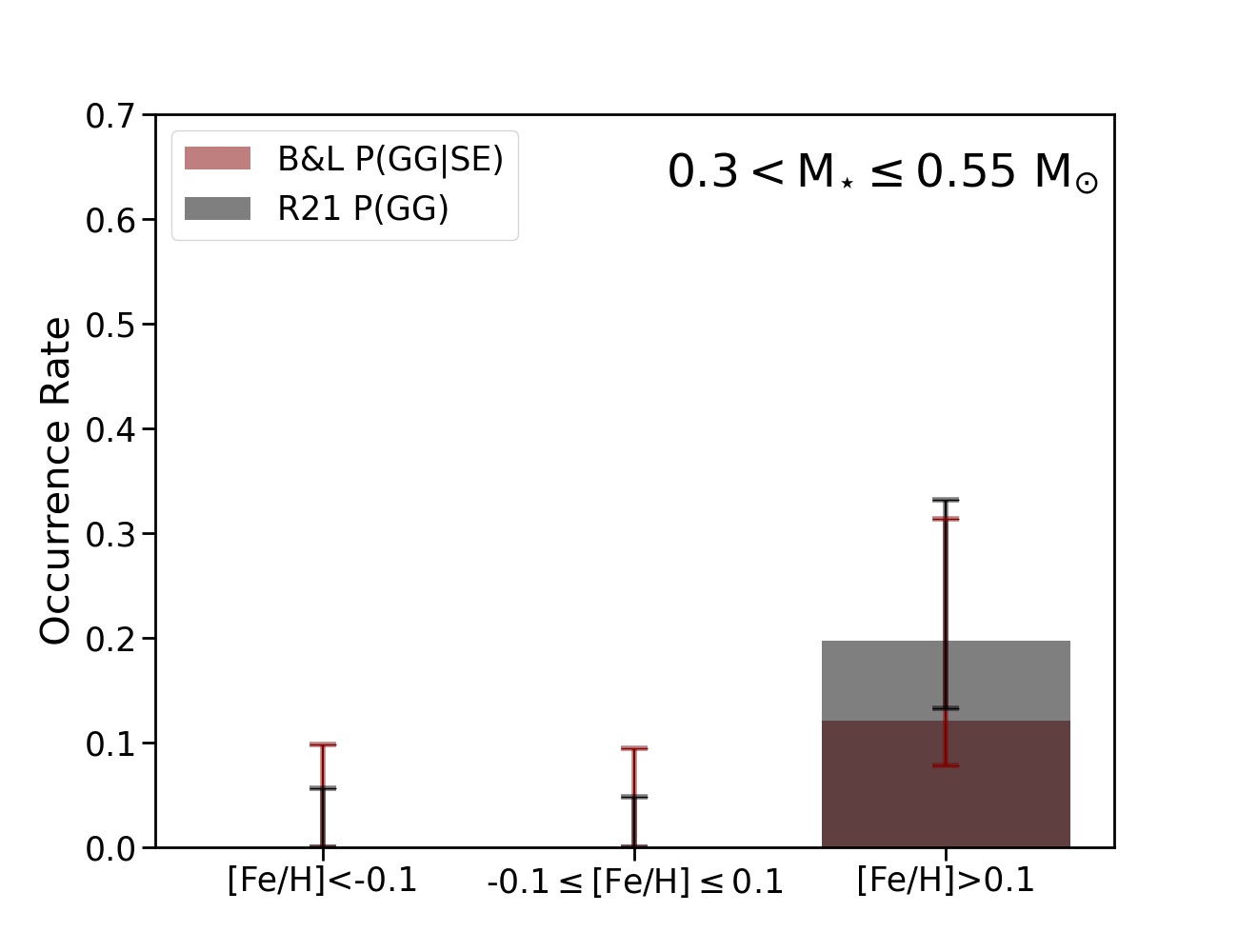} &
    \includegraphics[width=0.5\textwidth]{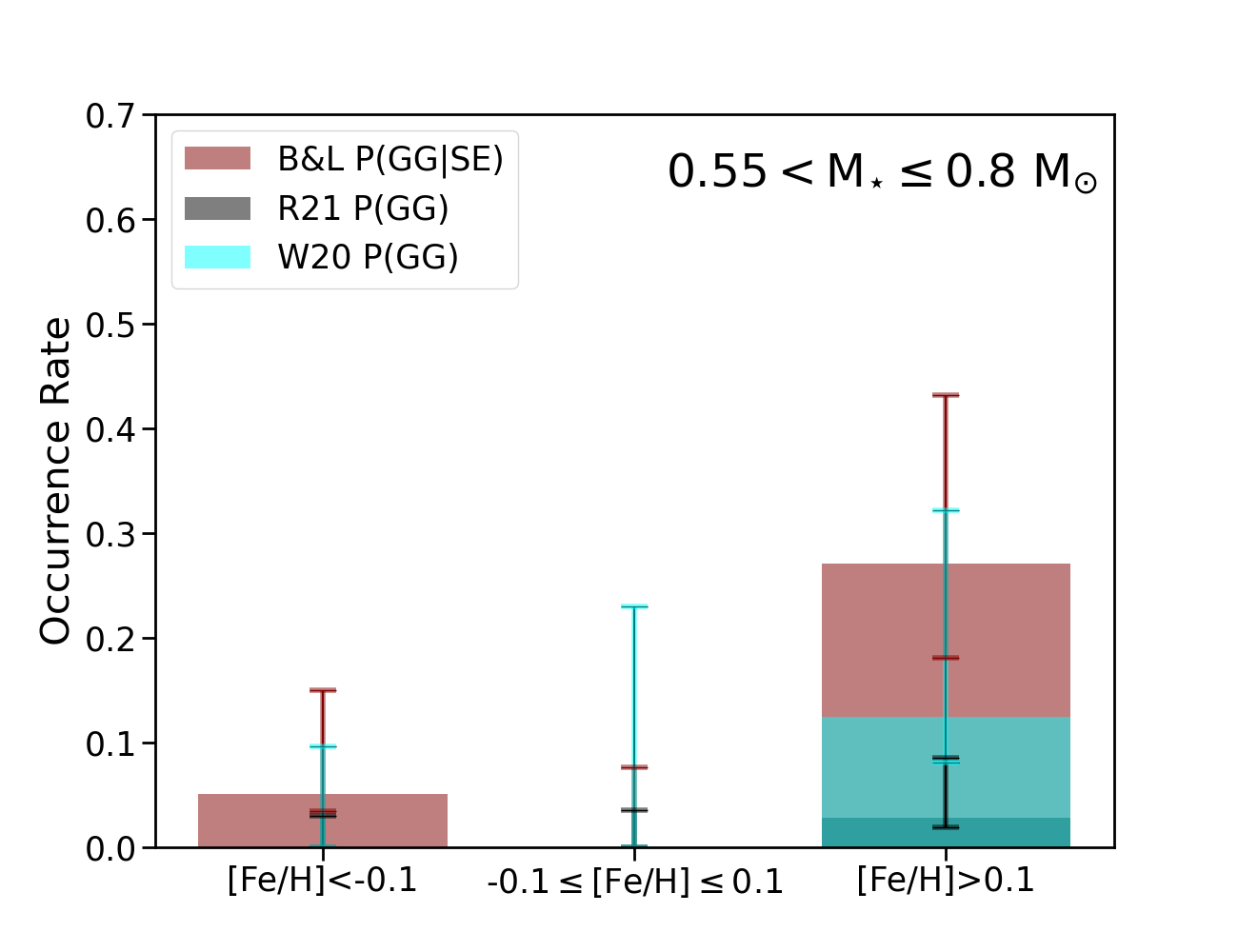}\\
    \includegraphics[width=0.5\textwidth]{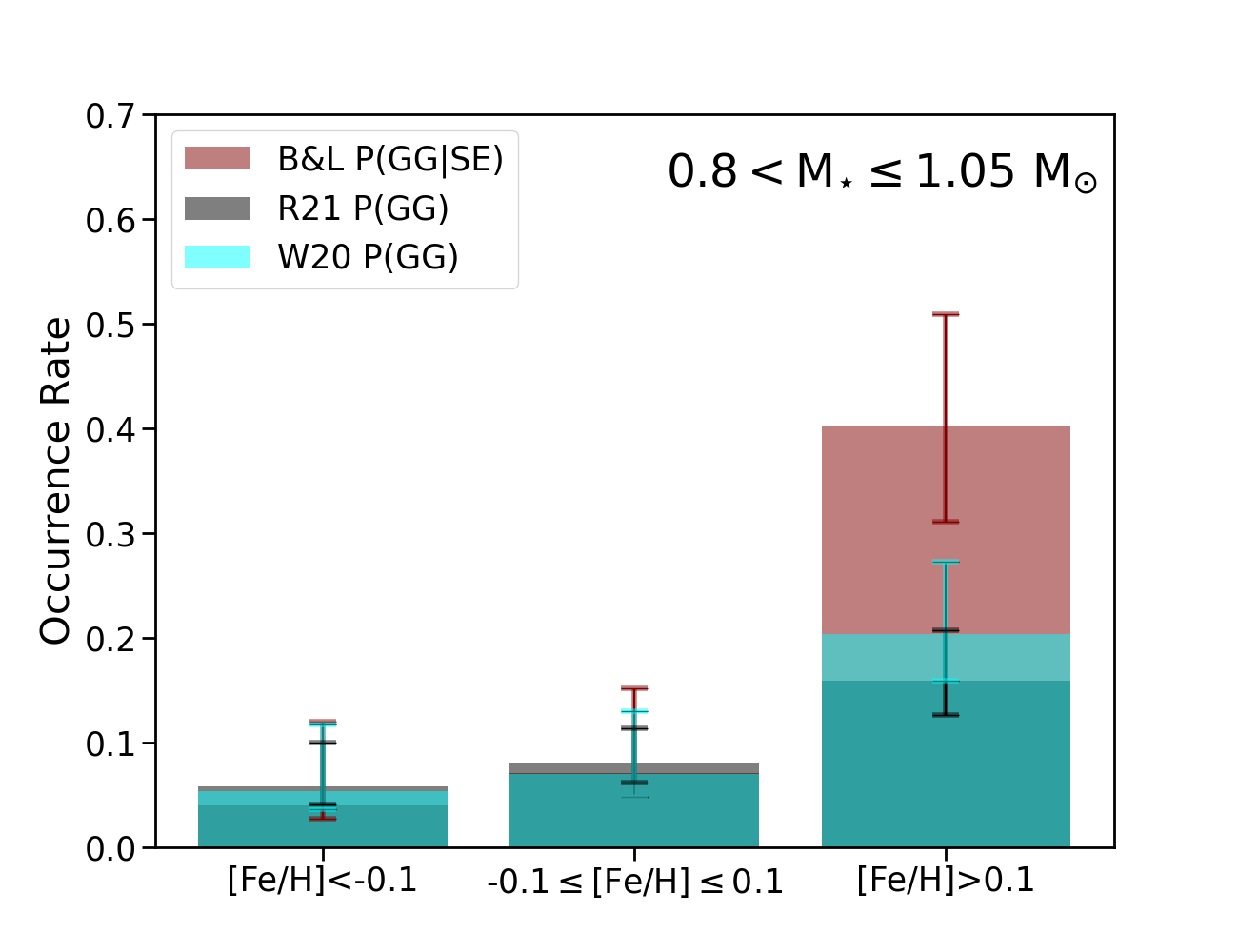} &
    \includegraphics[width=0.5\textwidth]{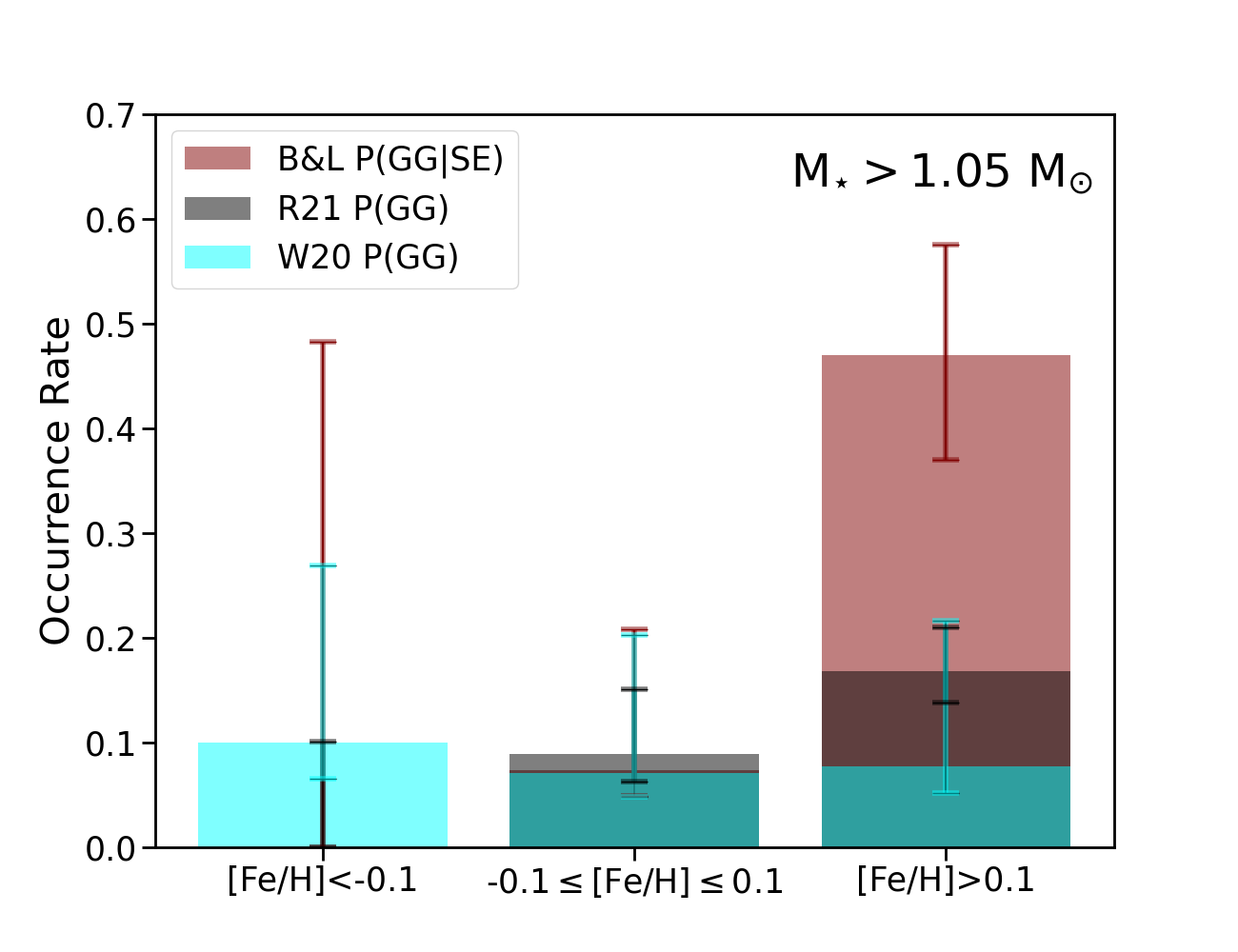}\\
    \end{tabular}
\caption{Comparison of gas giant (0.5--20 M$_{\rm Jup}$ 1--10 AU) frequencies from this paper's sample P(GG|SE), the R21 sample P(GG), and the W20 sample P(GG). Top Left: Comparison of low, middle, and high metallicity occurrence rates for early M-dwarfs. Top right: Same comparison for K-dwarfs. Bottom Left: Same comparison for G stars. Bottom right: Same comparison for F stars. For all panels, frequencies from this paper are shown in red, R21 in gray, and W20 in blue. We find a transition from no correlation between super-Earths and gas giants for M stars of any metallicity to a positive correlation in the highest metallicity bin. This transition occurs with K-dwarfs.}
\label{fig: resolve mass metallicity}
\end{figure*}

Figure \ref{fig: resolve mass metallicity} and Table \ref{table:occ-metal} show an evolution in the strength of the positive correlation between super-Earths and Jupiters through the stellar mass/metallicity parameter space. While even the most metal-rich M stars show no correlation between these planet populations, as the host stars get more massive the positive correlation in the most metal-rich bin appears and gets stronger. The transition between no correlation and a positive correlation at these high metallicities occurs in K-dwarf systems. No significant correlations are evident in the lower two metallicity bins [Fe/H]$<$0.1 at any of these stellar masses (see $\sigma$ significance values in Table \ref{table:occ-metal}). We note that the R21 and W20 occurrence rates are consistent to $<$0.6$\sigma$ for all mass/metallicity bins except one. The high metallicity K-dwarf occurrence rate for R21 is 1.3$\sigma$ lower than that of W20, P(GG) = 2.8$\%$ for R21 versus P(GG) = 12.5$\%$ for W20. Given the consistency of the gas giant occurrence rates P(GG) in the high metallicity bin for both samples ($\sim$15$\%$), we interpret the R21 P(GG) = 2.8$\%$ for the high metallicity K-dwarfs to be anomalously low, and the strength of the correlation shown in Table \ref{table:occ-metal} to be weaker than the quoted 2.3$\sigma$. Comparing to the W20 sample, the strength of the positive correlation for this subsample is 0.7$\sigma$, driven lower both by a higher frequency of gas giants in this bin and a smaller sample size with correspondingly larger error bars. 

\begin{deluxetable}{lccc}
\tabletypesize{\scriptsize}
\tablecaption{Gas Giant Occurrence Rates}
%\tablewidth{0pt}
\tablehead{
\colhead{M$_{\star}$ [M$_{\odot}$]} & \colhead{[Fe/H]$<$-0.1}& \colhead{-0.1$\leq$[Fe/H]$\leq$0.1}  & \colhead{[Fe/H]$>$0.1}
}
\startdata
0.3 -- 0.55 M$_{\odot}$ & &&\\
\hspace{0.1in} B$\&$L &$<$9.8$\%$& $<$9.5$\%$ & 12.1 (+19.2 -4.3)$\%$ \\
\hspace{0.1in} R21 & $<$5.7$\%$ & $<$4.8$\%$ & 19.7 (+13.5 -6.4)$\%$ \\
\hspace{0.1in} $\sigma$ & 0$\sigma$ & 0$\sigma$ & -0.4$\sigma$\\
0.55 -- 0.8 M$_{\odot}$ &&& \\
\hspace{0.1in} B$\&$L & 5.1 (+9.9 -1.6)$\%$ & $<$7.6$\%$ & 27.1 (+16.1 -9.0)$\%$ \\
\hspace{0.1in} R21 & $<$3.0$\%$ & $<$3.5$\%$ & 2.8 (+5.8 -1.0)$\%$ \tablenotemark{1}\\
\hspace{0.1in} $\sigma$ & 1.5$\sigma$ & 0$\sigma$ & 2.3$\sigma$\\
0.8 -- 1.05 M$_{\odot}$ &&& \\
\hspace{0.1in} B$\&$L & 4.0 (+8.0 -1.3)$\%$ & 7.1 (+8.1 -2.3)$\%$ & 40.2 (+10.7 -9.1)$\%$ \\
\hspace{0.1in} R21 & 5.8 (+4.1 -1.7)$\%$ & 8.1 (+3.3 -1.9)$\%$ & 15.9 (+4.8 -3.3)$\%$\\
\hspace{0.1in} $\sigma$ & -0.2$\sigma$ & -0.1$\sigma$ & 2.4$\sigma$ \\
$>$1.05 M$_{\odot}$ &&& \\
\hspace{0.1in} B$\&$L & $<$48.3$\%$ & 7.4 (+13.4 -2.5)$\%$ & 47.0 (+10.6 -10.0)$\%$\\
\hspace{0.1in} R21 & $<$10.1$\%$ & 8.9 (+6.2 -2.6)$\%$ & 16.8 (+4.2 -3.0)$\%$ \\
\hspace{0.1in} $\sigma$ & 0$\sigma$ & -0.1$\sigma$ & 2.8$\sigma$ \\
\enddata
\tablecomments{These occurrence rates use our default gas giant definition, namely 0.5--20 M$_{\rm Jup}$ and 1--10 AU.}
\tablenotetext{1}{This R21 occurrence rate is anomalously low, see section \ref{compare mstar sun} for details.}
\label{table:occ-metal}
\end{deluxetable}

\section{The Impact of Gas Giant Eccentricity, Separation, and Multiplicity}
\label{section: GG properties}

In BL24 we examined how the gas giant properties shaped the strength of the correlation in our FGK star sample. We found tentative evidence for a stronger positive correlation when the gas giant companions were farther away (outside 3AU), dynamically hot (eccentricity $>$0.2), and/or in multi-gas giant systems. Here we investigate how gas giant separation, eccentricity, and multiplicity shape the occurrence rates of gas giants as a function of host star mass and metallicity. We divide our combined sample into low and high metallicities ([Fe/H]$\leq$0 and [Fe/H]$>$0), and three mass bins (0.3--0.65 M$_{\odot}$, 0.65--1.0 M$_{\odot}$, and $>$1.0 M$_{\odot}$). Table \ref{table: occ rates} summarizes our results. 

We test the impact of gas giant separation by splitting gas giants into ones that are ``close'' (0.3--3AU), and ``distant'' (3--10AU). We use the inner bound of 0.3AU to match the tests of BL24 and \citet{Rosenthal2022}, though we note using an inner bound of 1AU does not significantly impact our conclusions. While there is no significant difference in the strength of the correlation for the two lowest mass bins of metal-rich stars (M$_{\star}$$<$1.0 M$_{\odot}$), the positive correlation is stronger for more distant gas giants around G and F stars. 

Next, we consider the impact of dynamical activity of the outer giants, dividing our sample into systems with dynamically hot gas giants ($e>0.2$), and dynamically cold gas giants ($e<0.2$). For all three high-metallicity stellar mass bins, we see that the dynamically hot systems produce a stronger positive correlation than the dynamically cold systems. 

Finally, we examine how gas giant multiplicity might shape the correlation. For the lowest mass stars, there is no correlation for systems with either one or more gas giants. While the high-metallicity middle mass bin shows a tentative strengthening of the positive correlation when there is a single gas giant in the system, this trend flips for the highest mass stars, with multi-gas giant systems strengthening the positive correlation over single gas giant systems. 

\begin{deluxetable*}{lcccccc}
\tabletypesize{\scriptsize}
\tablecaption{Occurrence Rates GG Properties}
%\tablewidth{0pt}
\tablehead{
\colhead{M$_{\star}$ [M$_{\odot}$]} & \colhead{B$\&$L[Fe/H]$\leq$0}& \colhead{R21 [Fe/H]$\leq$0}& \colhead{B$\&$L [Fe/H]$>$0}& \colhead{R21 [Fe/H]$>$0} & \colhead{$\sigma$[Fe/H]$\leq$0}& \colhead{$\sigma$[Fe/H]$>$0}
}
\startdata
0.3 -- 0.65 M$_{\odot}$ & &&\\
\hspace{0.1in} Close GG (0.3-3AU)&$<$4.2$\%$& $<$2.5$\%$ & 10.5 (+11.1 -3.5)$\%$ & 8.5 (+5.9 -2.5)$\%$ & 0$\sigma$& 0.3$\sigma$\\
\hspace{0.1in} Distant GG (3-10AU)&$<$4.5$\%$& $<$2.5$\%$ & 5.7 (+11.0 -1.8)$\%$ & 4.3 (+5.2 -1.4)$\%$ & 0$\sigma$& 0.3$\sigma$ \\
\hspace{0.1in} Dynamically Hot&$<$4.4$\%$& $<$2.5$\%$ & 5.5 (+10.7 -1.7)$\%$ & 2.2 (+4.6 -0.7)$\%$ & 0$\sigma$& 0.7$\sigma$\\
\hspace{0.1in} Dynamically Cold&$<$4.4$\%$& $<$2.5$\%$ & 5.5 (+10.7 -1.7)$\%$ & 8.6 (+6.0 -2.5)$\%$ & 0$\sigma$& -0.3$\sigma$\\
\hspace{0.1in} Multi GG&$<$4.1$\%$& $<$2.9$\%$ & 5.2 (+10.1 -1.7)$\%$ & 4.3 (+5.1 -1.4)$\%$ & 0$\sigma$& 0.2$\sigma$\\
\hspace{0.1in} Single GG&$<$4.1$\%$& $<$2.5$\%$ & 5.2 (+10.1 -1.7)$\%$ & 4.3 (+5.1 -1.4)$\%$  & 0$\sigma$& 0.2$
\sigma$\\
0.65 -- 1.0 M$_{\odot}$ &&& \\
\hspace{0.1in} Close GG (0.3-3 AU)& $<$2.9$\%$ & 2.4 (+1.8 -0.7)$\%$ & 13.4 (+6.6 -3.6)$\%$ & 5.2 (+2.4 -1.3)$\%$ & -0.8$\sigma$& 1.9$\sigma$\\
\hspace{0.1in} Distant GG (3-10AU)& 1.8 (+3.8 -0.6)$\%$& 1.8 (+1.7 -0.5)$\%$ & 12.0 (+6.8 -3.3)$\%$ & 4.6 (+2.3 -1.2)$\%$ & 0$\sigma$& 1.8$\sigma$\\
\hspace{0.1in} Dynamically Hot& $<$3.0$\%$& $<$1.1$\%$ & 16.3 (+7.1 -4.1)$\%$ & 6.5 (+2.6 -1.4)$\%$ & 0$\sigma$& 2.0 $\sigma$\\
\hspace{0.1in} Dynamically Cold& 1.7 (+3.7 -0.5)$\%$& 3.6 (+2.0 -1.0)$\%$ & 4.7 (+5.5 -1.6)$\%$ & 3.3 (+2.1 -0.9)$\%$  & -0.5$\sigma$& 0.5$\sigma$\\
\hspace{0.1in} Multi GG&$<$2.9$\%$& 0.6 (+1.3 -0.2)$\%$ & 4.4 (+5.3 -1.4)$\%$ & 1.9 (+1.9 -0.6)$\%$ & -0.2$\sigma$& 1.0$\sigma$\\
\hspace{0.1in} Single GG& 1.6 (+3.5 -0.5)$\%$& 2.9 (+2.0 -0.8)$\%$ & 15.4 (+6.8 -3.9)$\%$ & 6.5 (+2.5 -1.5)$\%$  & -0.4$\sigma$& 1.9$\sigma$\\
$>$1.0 M$_{\odot}$ &&& \\
\hspace{0.1in} Close GG (0.3-3 AU) & 6.3 (+11.8 -2.1)$\%$ & 6.0 (+5.3 -1.8)$\%$ & 22.6 (+7.8 -5.2)$\%$ & 13.6 (+3.0 -2.2)$\%$ & 0.1$\sigma$ & 1.5$\sigma$\\
\hspace{0.1in} Distant GG (3-10AU)&$<$11.3$\%$& 4.1 (+5.0 -1.3)$\%$ & 24.5 (+8.3 -5.6)$\%$ & 9.6 (+2.8 -1.8)$\%$ & -0.4$\sigma$ & 2.4$\sigma$\\
\hspace{0.1in} Dynamically Hot& 6.7 (+12.5 -2.2)$\%$& 6.1 (+5.3 -1.9)$\%$ & 23.7 (+8.1 -5.4)$\%$ & 8.4 (+2.6 -1.7)$\%$ & 0.1$\sigma$ & 2.6$\sigma$\\
\hspace{0.1in} Dynamically Cold&$<$10.8$\%$& $<$3.6$\%$ & 21.1 (+8.0 -5.1)$\%$ & 10.7 (+2.9 -1.9)$\%$ & 0$\sigma$& 1.8$\sigma$\\
\hspace{0.1in} Multi GG&$<$10.1$\%$& $<$3.5$\%$ & 9.9 (+6.7 -2.9)$\%$ & 2.4 (+1.7 -0.7)$\%$ & 0$\sigma$& 2.2$\sigma$\\
\hspace{0.1in} Single GG& 6.2 (+11.7 -2.0)$\%$& 6.0 (+5.3 -1.8)$\%$ & 24.8 (+7.8 -5.5)$\%$ & 15.3 (+3.2 -2.3)$\%$ & 0.1$\sigma$& 1.5$\sigma$\\
\enddata
\tablecomments{Default: GG 1--10 AU, 0.5--20 M$_{\rm Jup}$; Close GG: 0.3--3 AU, 0.5--20 M$_{\rm Jup}$; Distant GG:  3--10 AU, 0.5--20 M$_{\rm Jup}$; Dynamically Hot: default GG with eccentricities $>$0.2; Dynamically Cold: default GG with eccentricities $<$0.2; Single GG: systems with a single gas giant in the default GG range; Multi GG: systems with multiple gas giants in the default GG range.}
\label{table: occ rates}
\end{deluxetable*}

\section{Discussion and Conclusions}
\label{section: discussion}

\begin{figure}
    \centering
    \includegraphics[width=0.5\textwidth]{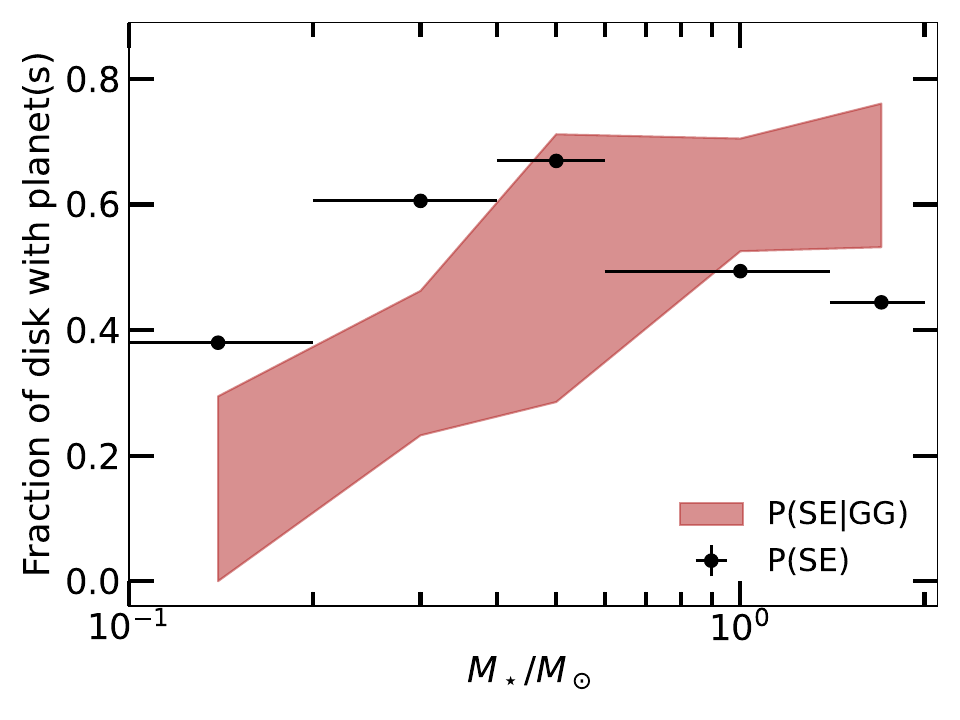}
    \caption{Fraction of disks expected to create inner (inside 100 days) super-Earths (black) compared to the fraction of disks with enough mass to create both the outer giant and inner super-Earths bounded by thrice as much (lower edge of the red zone) and twice as much (upper edge of the red zone) required to create a single giant at 2000 days. See text for more discussion on our definition. The span of the horizontal lines for each black circles illustrate the range of stellar masses used for each bin in the calculation of disk fraction. We use the methodology of \citet{Chachan23} and we choose turbulent $\alpha_{\rm t}=10^{-3}$, $\dot{M}_{\star,\odot}=10^{-8}M_\odot\,{\rm yr}^{-1}$, and fragmentation velocity of 10 m/s as a representative illustration (we obtain qualitatively similar results at higher fragmentation velocities).}
    \label{fig:Psegg}
\end{figure}

The key result of this paper is the disappearance of inner-outer planet correlation at stellar masses below $\sim$0.55$M_\odot$, even when we limit our sample to high metallicity hosts. We close our writing with a discussion on the potential origin of this feature.

First, we consider whether there is any evidence of disks being more compact around low mass stars and thereby limiting the spatial extent over which planets can form even if there was enough mass to create a giant near its outer edge. The median and standard deviation orbital distance of outer giants (1--10 AU, 0.5--20 M$_{\rm Jup}$) in our sample is 2.52, 2.44$\pm$3.3, 3.17$\pm$2.0, and 3.24$\pm$1.5 AU for our four stellar mass bins, from lightest to heaviest. Number of gas giants in each bin are 1, 6, 15, and 13 respectively. We therefore do not see strong evidence of the disk being primordially more compact around low mass stars, at least insofar as how it bears on the currently observed properties of the outer giant. From the observations of protoplanetary disks, it is difficult to verify whether the disks are preferentially more compact around low mass stars. While there is a positive correlation between the stellar mass and mm luminosity \citep[e.g.,][]{Andrews13} and a positive correlation between mm luminosity and the size of the disk \citep[e.g.,][]{Tripathi17}, the latter could simply be an effect of local dust trapping in the outer orbits. Whether the incidence of such substructures have any dependence on stellar mass is unclear and recent observational reports are mixed \citep{VanDerMarel21,Shi24}. 

Next, we consider the relationship between total disk mass budget and the mass of the host star and how this relationship bears on the ability of the disk to convert solids into planets.
Based on simple consideration of pebble accretion, \citet{Chachan23} hypothesized that the positive correlation between the inner super-Earths and outer gas giants should persist and may even strengthen for lower mass stars. Our result in this paper appears to be in direct contradiction. The statement of \citet{Chachan23} is based on the mass required to create the outer giants $M_{\rm out}$ being larger than that required to create the inner planets $M_{\rm in}$ (and also for the coagulation time of inner cores to be shorter than the radial drift time, which is ensured for similar fragmentation velocities of silicate and ice-coated grains; e.g., \citealt{Kimura20}). This is a sufficient condition if the total disk mass $M_{\rm disk}$ is significantly greater than $M_{\rm out}$ but as $M_{\rm disk}$ approaches that of $M_{\rm out}$, the sufficiency diminishes, and $M_{\rm disk}$ is seen to generally decrease around lower mass stars \citep{Manara23}.

We now consider how the lack of positive correlation in P(GG|SE) can be understood under the framework outlined in \citet{Chachan23}. In this paper, we define a positive inner-outer planet correlation as follows:

\begin{equation}
    \rm P(GG|SE) = \frac{P(GG)\times P(SE|GG)}{P(SE)} > P(GG),
\end{equation}
which is equivalent to
\begin{equation}
    \rm P(SE) < P(SE|GG).
    \label{eq:corr-conditon}
\end{equation}
Over a large swath of the parameter space explored in \citet{Chachan23}, the fraction of systems harboring inner super-Earths P(SE) rises towards low mass stars owing to the higher pebble accretion efficiency but only down to $\sim$0.5$M_\odot$.\footnote{The pebble accretion efficiency is defined as the ratio of the pebble accretion rate to the radial drift rate of the pebbles. This efficiency is higher towards low mass stars because of the shallower gravitational potential of the star---which boosts the effective gravity of the accreting body and therefore its pebble accretion rate---and the disk is colder, leading to denser disk again boosting the pebble accretion rate while also leading to slower headwind slowing down the radial drift of pebbles.} Below this mass, P(SE) declines towards lighter stars as their protoplanetary disks often do not have enough mass to create even the super-Earths.\footnote{\citet{Mulders21} arrive at similar result of decreasing P(SE) towards lighter stars below $\sim$0.5$M_\odot$ but they predict anti-correlation between inner super-Earths and outer giants because of their assumption of significantly different fragmentation velocity between silicate and ice-coated grains. Such an assumption has been called into question by modern laboratory measurements \citep[e.g.,][]{Musiolik19,Kimura20}.} 
We estimate P(SE|GG) as the fraction of disks with 2--3$\times$ the mass required to nucleate a core of 15$M_\oplus$ at an orbital period of 2000 days. We choose 15$M_\oplus$ to ensure runaway gas accretion within $\sim$1 Myr to safely create a gas giant. This critical core mass depends on the opacity of the accreted material and can be lower in low opacity environment or higher in high opacity environment \citep[e.g.,][]{Ikoma00,Lee14,Piso15,Savignac24}. Our choice of using 2--3$\times$ the required mass for a single giant to define P(SE|GG) ensures the disk has enough mass to create 1--2 outer giants and multiple inner super-Earths across the range of stellar masses. 

Figure \ref{fig:Psegg} demonstrates the inequality \ref{eq:corr-conditon} to hold down to $M_\star \sim 0.5M_\odot$ below which $P(SE) > P(SE|GG)$, consistent with the observed lack of correlation around low mass host stars. The shape of P(SE|GG) vs.~stellar mass is driven by the need for high enough disk mass reservoir to generate a planetary system with both inner super-Earths and outer gas giants and the smaller fraction of such massive disks around lower mass stars. We note that our definition of P(SE|GG) is approximate. For instance, using 10$M_\oplus$ as the critical core mass will slightly shift P(SE|GG) up so that the upper edge of the P(SE|GG) exceeds P(SE) at the lowest stellar mass. Even so, our qualitative result stays the same in that the inequality \ref{eq:corr-conditon} is assured at stellar masses $\gtrsim$0.6$M_\odot$, but the situation flips at lower stellar masses so that the inner-outer planet correlation weakens or disappears. 

The same effect of disk mass budget may also explain the switch in the strength of the inner-outer planet correlation towards single to multiple gas giants from low (0.65--1.0$M_\odot$) to high mass ($>1.0M_\odot$) stars. Around lower mass stars, disks are generally lighter so that they may run out of all the mass if they ended up creating multiple giants. By contrast, around high mass stars, disks are generally heavier so that they may have more than enough mass to create multiple giants and also the inner super-Earths.

A larger sample of disk masses over the full range of stellar mass and more detailed calculation of P(SE|GG) marginalized over the full parameter space would help refining the quantitative details of our calculations, which we defer to a future study.

\vspace{0.5cm}
M.L.B. acknowledges support by NSERC, the Heising-Simons Foundation, and the Connaught New Researcher Award from the University of Toronto.  E.J.L. gratefully acknowledges support by NSERC, by FRQNT, by the Trottier Space Institute, and by the William Dawson Scholarship from McGill University.

\appendix
\counterwithin{figure}{section}
\counterwithin{table}{section}

\section{Full M-dwarf sample list}

Below we show our sample of M-dwarf super-Earth host stars as well as average sensitivity maps for our sample and the R21 comparison sample.

\startlongtable
\begin{deluxetable*}{lcccccccc}
\tabletypesize{\scriptsize}
\tablecaption{Sample of systems}
\tablewidth{0pt}
\tablehead{
\colhead{Target} & \colhead{M$_{\star}$ (M$_{\odot}$)} & \colhead{[Fe/H]} & \colhead{N$_{\rm pl}$} & \colhead{Disc. Method} & \colhead{N$_{obs}$} & \colhead{Baseline (yrs)} & \colhead{GG Companion}  & \colhead{Ref.}
}
\startdata
AU Mic & 0.50$\pm$0.03 & 0.12$\pm$0.10 & 3 & Transit & 431 & 17.0 & No & 65\\
CD Cet & 0.16$\pm$0.01 & 0.13$\pm$0.16 & 1 & RV & 228 & 3.2 & No & 1\\
G 264-012 & 0.30$\pm$0.02 & 0.10$\pm$0.19 & 2 & RV & 305 & 3.6 & No & 2 \\
GJ 12 & 0.24$\pm$0.01 & -0.30$\pm$0.10 & 1 & Transit & 93 & 4.2 & No & 67\\
GJ 15A & 0.38$\pm$0.05 & -0.34$\pm$0.09 & 2 & RV & 439 & 22.8 & No & 7\\
GJ 27.1 & 0.53$\pm$0.10\tablenotemark{2} & -0.09$\pm$0.10\tablenotemark{1} & 1 & RV & 50 & 1.2 & No & 10\\
GJ 49 & 0.52$\pm$0.02 & 0.13$\pm$0.16 & 1 & RV & 238 & 20.6 & No & 35\\
GJ 163 & 0.40$\pm$0.02 & -0.02$\pm$0.10\tablenotemark{1} & 3 & RV & 153 & 8.4 & No & 64\\
GJ 229 A & 0.51 $\pm$0.02 & 0.15$\pm$0.10\tablenotemark{1} & 2 & RV & 531 & 21.0 & No & 8 \\
GJ 251 & 0.35$\pm$0.02 & -0.03$\pm$0.16 & 1 & RV & 212 & 4.0 & No & 9\\
GJ 273 & 0.29$\pm$0.10\tablenotemark{2} & 0.09$\pm$0.17 & 2 & RV & 280 & 12.8 & No & 11\\
GJ 357 & 0.34$\pm$0.01 & -0.12$\pm$0.16 & 3 & Transit & 141 & 21.3 & No & 14 \\
GJ 367 & 0.46$\pm$0.01 & -0.01$\pm$0.12 & 3 & Transit & 370 & 2.8 & No & 16\\
GJ 378 & 0.56$\pm$0.01 & 0.06$\pm$0.09 & 1 & RV & 44 & 2.4 & No & 34\\
GJ 393 & 0.43$\pm$0.02 & -0.09$\pm$0.16 & 1 & RV & 334 & 23.0 & No & 2\\
GJ 411 & 0.39$\pm$0.01 & -0.36 (+0.09 -0.07) & 2 & RV & 704 & 32.9 & No & 7\\
GJ 422 & 0.35$\pm$0.10\tablenotemark{2} & 0.18$\pm$0.10\tablenotemark{1} & 1 & RV & 128 & 14.2 & No & 8\\
GJ 433 & 0.48$\pm$0.10\tablenotemark{2} & -0.22$\pm$0.10\tablenotemark{1} & 3 & RV & 415 & 20.1 & No & 8\\
GJ 480 & 0.45$\pm$0.02 & -0.06$\pm$0.10\tablenotemark{1} & 1 & RV & 21 & 4.7 & No & 21\\
GJ 486 & 0.32$\pm$0.02 & 0.07$\pm$0.16 & 1 & Transit & 170 & 22.4 & No & 22\\
GJ 514 & 0.51$\pm$0.05 & -0.14$\pm$0.09 & 1 & RV & 436 & 16.9 & No & 23\\
GJ 536 & 0.52$\pm$0.05 & -0.08$\pm$0.09 & 1 & RV & 158 & 11.9 & No & 24\\
GJ 581 & 0.31$\pm$0.02 & -0.09$\pm$0.07 & 3 & RV & 446 & 19.4 & No & 7\\
GJ 625 & 0.30$\pm$0.07 & -0.38$\pm$0.09 & 1 & RV & 140 & 3.3 & No & 25\\
GJ 667 C & 0.33$\pm$0.02 & -0.55$\pm$0.10 & 2 & RV & 173 & 7.8 & No & 26\\
GJ 674 & 0.35$\pm$0.10\tablenotemark{2} & -0.28$\pm$0.10\tablenotemark{1} & 1 & RV & 32 & 2.3 & No & 27\\
GJ 685 & 0.55$\pm$0.06 & -0.15$\pm$0.09 & 1 & RV & 106 & 4.4 & No & 28\\
GJ 686 & 0.43$\pm$0.02 & -0.23$\pm$0.16 & 1 & RV & 172 & 12.8 & No & 36\\
GJ 687 & 0.42$\pm$0.01 & 0.06$\pm$0.08 & 2 & RV & 160 & 16.3 & No & 7\\
GJ 720 A & 0.57$\pm$0.06 & -0.14$\pm$0.09 & 1 & RV & 132 & 7.2 & No & 29\\
GJ 724 & 0.53$\pm$0.03 & -0.02$\pm$0.05 & 1 & RV & 110 & 14.4 & No & 18\\
GJ 740 & 0.58$\pm$0.06 & 0.08$\pm$0.16 & 1 & RV & 201 & 11.0 & No & 30\\
GJ 806 & 0.41$\pm$0.01 & -0.28$\pm$0.0.07 & 2 & Transit & 67 & 5.5 & No & 31\\
GJ 876 & 0.37$\pm$0.01 & 0.21$\pm$0.06 & 4 & RV & 354 & 23.5 & Yes & 7\\
GJ 887 & 0.49$\pm$0.05 & -0.06$\pm$0.08 & 2 & RV & 288 & 20.2 & No & 32\\
GJ 1002 & 0.12$\pm$0.01 & -0.25$\pm$0.19 & 2 & RV & 139 & 5.4 & No & 3 \\
GJ 1132 & 0.18$\pm$0.02 & -0.12$\pm$0.15 & 2 & Transit & 138 & 2.0 & No & 4\\
GJ 1151 & 0.16$\pm$0.01 & -0.12$\pm$0.10 & 1 & RV & 145 & 6.3 & No & 5\\
GJ 1214 & 0.18$\pm$0.01 & 0.24$\pm$0.11 & 1 & Transit & 165 & 10.2 & No & 49\\
GJ 1265 & 0.18$\pm$0.02 & -0.04$\pm$0.16 & 1 & RV & 98 & 11.4 & No & 6\\
GJ 3082 & 0.47$\pm$0.10\tablenotemark{2} & -0.22$\pm$0.10\tablenotemark{1} & 1 & RV & 94 & 7.4 & No & 8\\
GJ 3293 & 0.42$\pm$0.10\tablenotemark{2} & 0.02$\pm$0.09 & 4 & RV & 207 & 6.3 & No & 11\\
GJ 3323 & 0.16$\pm$0.10\tablenotemark{2} & -0.27$\pm$0.09 & 2 & RV & 154 & 11.9 & No & 11\\
GJ 3341 & 0.47$\pm$0.10\tablenotemark{2} & -0.09$\pm$0.09 & 1 & RV & 135 & 4.0 & No & 12\\
GJ 3470 & 0.54 (+0.05 -0.04) & 0.20$\pm$0.10 & 1 & RV & 108 & 5.1 & No & 13\\
GJ 3634 & 0.45$\pm$0.05 & -0.05$\pm$0.10\tablenotemark{1} & 1 & RV & 54 & 1.3 & No & 15\\
GJ 3779 & 0.27$\pm$0.02 & 0.0$\pm$0.20 & 1 & RV & 104 & 2.3 & No & 6\\
GJ 3929 & 0.31 (+0.03 -0.02) & -0.02$\pm$0.12 & 2 & Transit & 42 & 1.5 & No & 17\\
GJ 3988 & 0.18$\pm$0.02 & -0.12$\pm$0.17 & 1 & RV & 164 & 6.3 & No & 18\\
GJ 3998 & 0.50$\pm$0.05 & -0.16$\pm$0.09 & 2 & RV & 136 & 2.4 & No & 19\\
GJ 4276 & 0.41$\pm$0.03 & 0.12$\pm$0.16 & 1 & RV & 100 & 2.1 & No & 20\\
GJ 9689 & 0.59$\pm$0.06 & 0.05$\pm$0.04 & 1 & RV & 174 & 7.4 & No & 33\\
HD 33793 & 0.28$\pm$0.01 & -0.89$\pm$0.10\tablenotemark{1} & 1 & RV & 135 & 10.3 & No & 39\\
HD 180617 & 0.48$\pm$0.02 & -0.04$\pm$0.16 & 1 & RV & 274 & 15.2 & No & 36\\
HD 238090 & 0.58$\pm$0.02 & -0.03$\pm$0.16 & 1 & RV & 108 & 3.3 & No & 9\\
HD 285968 & 0.51$\pm$0.01 & 0.15$\pm$0.08 & 1 & RV & 121 & 21.6 & No & 7\\
HIP 22627 & 0.36$\pm$0.03 & 0.30$\pm$0.10 & 2 & RV & 78 & 19.9 & Yes & 7\\
HIP 54373 & 0.57$\pm$0.03 & 0.03$\pm$0.10\tablenotemark{1} & 2 & RV & 51 & 8.2 & No & 37\\
HIP 83043 & 0.51$\pm$0.05 & -0.15$\pm$0.09 & 2 & RV & 81 & 14.5 & No & 7\\
HN Lib & 0.29$\pm$0.01 & -0.18$\pm$0.15 & 1 & RV & 101 & 4.9 & No & 38\\
K2-3 & 0.55$\pm$0.03 & -0.16$\pm$0.08 & 3 & Transit & 195 & 2.5 & No & 52\\
K2-18 & 0.36$\pm$0.05 & 0.12$\pm$0.16 & 1 & Transit & 75 & 2.1 & No & 50\\
K2-25 & 0.26$\pm$0.01 & 0.15$\pm$0.03 & 1 & Transit & 63 & 1.2 & No & 51\\
K2-415 & 0.16$\pm$0.01 & -0.13$\pm$0.18 & 1 & Transit & 42 & 3.4 & No & 53\\
Kepler-138 & 0.54$\pm$0.01 & -0.18$\pm$0.10 & 4 & Transit & 29 & 4.4 & No & 54\\
L 98-59 & 0.27$\pm$0.03 & -0.46$\pm$0.26 & 4 & Transit & 66 & 1.4 & No & 41\\
L 363-38 & 0.21$\pm$0.01 & -0.16$\pm$0.10\tablenotemark{1} & 1 & RV & 31 & 1.2 & No & 40\\
LHS 1140 & 0.18$\pm$0.01 & -0.15$\pm$0.09 & 2 & Transit & 293 & 2.1 & No & 55\\
LHS 1815 & 0.50$\pm$0.02 & -0.12$\pm$0.09 & 1 & Transit & 86 & 16.3 & No & 56\\
LSPM J2116+0234 & 0.43$\pm$0.03 & -0.05$\pm$0.16 & 1 & RV & 125 & 2.4 & No & 42\\
LTT 1445A & 0.26$\pm$0.01 & -0.34$\pm$0.09 & 2 & Transit & 136 & 1.8 & No & 57\\
Proxima Cen & 0.12$\pm$0.01 & -0.04$\pm$0.10\tablenotemark{1} & 1 & RV & 114 & 2.2 & No & 43\\
Ross 128 & 0.17$\pm$0.02 & -0.02$\pm$0.08 & 1 & RV & 159 & 10.8 & No & 44\\
Ross 508 & 0.18$\pm$0.01 & -0.20$\pm$0.20 & 1 & RV & 102 & 2.6 & No & 45\\
Teegarden's Star & 0.10$\pm$0.01 & -0.11$\pm$0.28 & 3 & RV & 467 & 5.7 & No & 46\\
TOI-244 & 0.43$\pm$0.03 & -0.39$\pm$0.07 & 1 & Transit & 72 & 3.8 & No & 63\\
TOI-663 & 0.51$\pm$0.01 & 0.07$\pm$0.12 & 3 & Transit & 83 & 2.1 & No & 66\\
TOI-1266 & 0.43$\pm$0.02 & -0.20$\pm$0.12 & 2 & Transit & 146 & 2.9 & No & 58\\
TOI-1452 & 0.25$\pm$0.01 & -0.07$\pm$0.02 & 1 & Transit & 215 & 1.1 & No & 59\\
TOI-1468 & 0.34$\pm$0.01 & -0.04$\pm$0.07 & 2 & Transit & 97 & 1.9 & No & 60\\
TOI-1695 & 0.51$\pm$0.01 & 0$\pm$0.10\tablenotemark{1} & 1 & Transit & 46 & 1.1 & No & 61\\
TOI-2018 & 0.57$\pm$0.02 & -0.58$\pm$0.18 & 1 & Transit & 38 & 10.2 & No & 62\\
Wolf 1061 & 0.29$\pm$0.10\tablenotemark{2} & -0.09$\pm$0.09 & 3 & RV & 187 & 11.3 & 11\\
Wolf 1069 & 0.17$\pm$0.01 & 0.07$\pm$0.19 & 1 & RV & 262 & 4.0 & 47\\
YZ Cet & 0.14$\pm$0.01 & -0.18$\pm$0.16 & 3 & RV & 440 & 15.1 & No & 48
\enddata
\tablecomments{References: (1) \citet{Bauer2020}, (2) \citet{Amado2021}, (3) \citet{Suarez2023}, (4) \citet{Bonfils2018}, (5) \citet{Blanco2023}, (6) \citet{Luque2018}, (7) \citet{Rosenthal2021}, (8) \citet{Feng2020}, (9) \citet{Stock2020}, (10) \citet{Tuomi2014}, (11) \citet{Astudillo2017}, (12) \citet{Astudillo2015}, (13) \citet{Kosiarek2019}, (14) \citet{Luque2019}, (15) \citet{Bonfils2011}, (16) \citet{Goffo2023}, (17) \citet{Beard2022}, (18) \citet{Gorrini2023}, (19) \citet{Affer2016}, (20) \citet{Nagel2019}, (21) \citet{Feng2020_2}, (22) \citet{Trifonov2021}, (23) \citet{Damasso2022}, (24) \citet{Suarez2017}, (25) \citet{Suarez2017_2}, (26) \citet{Anglada2013}, (27) \citet{Bonfils2007}, (28) \citet{Pinamonti2019}, (29) \citet{Gonzalez2021}, (30) \citet{Toledo2021}, (31) \citet{Palle2023}, (32) \citet{Jeffers2020}, (33) \citet{Maldonado2021}, (34) \citet{Hobson2019}, (35) \citet{Perger2019}, (36) \citet{Burt2021}, (37) \citet{Feng2019}, (38) \citet{Gonzalez2023}, (39) \citet{Anglada2014}, (40) \citet{Sartori2023}, (41) \citet{Demangeon2021}, (42) \citet{Lalitha2019}, (43) \citet{Faria2022}, (44) \citet{Bonfils2018_2}, (45) \citet{Harakawa2022}, (46) \citet{Dreizler2024}, (47) \citet{Kossakowski2023}, (48) \citet{Stock2020_2}, (49) \citet{Cloutier2021}, (50) \citet{Cloutier2017}, (51) \citet{Stefansson2020}, (52) \citet{Bonomo2023}, (53) \citet{Hirano2023}, (54) \citet{Piaulet2023}, (55) \citet{Ment2019}, (56) \citet{Luque2022}, (57) \citet{Winters2022}, (58) \citet{Cloutier2024}, (59) \citet{Cadieux2022}, (60) \citet{Chaturvedi2022}, (61) \citet{Cherubim2023}, (62) \citet{Dai2023}, (63) \citet{Castro2023}, (64) \citet{Bonfils2013}, (66) \citet{Cointepas2024}, (67) \citet{Kuzuhara2024}}
\tablenotetext{1}{In the cases where uncertainties on stellar metallicities are not given, we assume an error of 0.1 dex.}
\tablenotetext{2}{In the cases where uncertainties on stellar mass are not given, we assume an error of 0.1 M$_{\sun}$}
\label{tab:full-sample}
\end{deluxetable*}

\begin{figure}
 %   \centering
%    \begin{tabular}{cc}
    \includegraphics[width=0.5\textwidth]{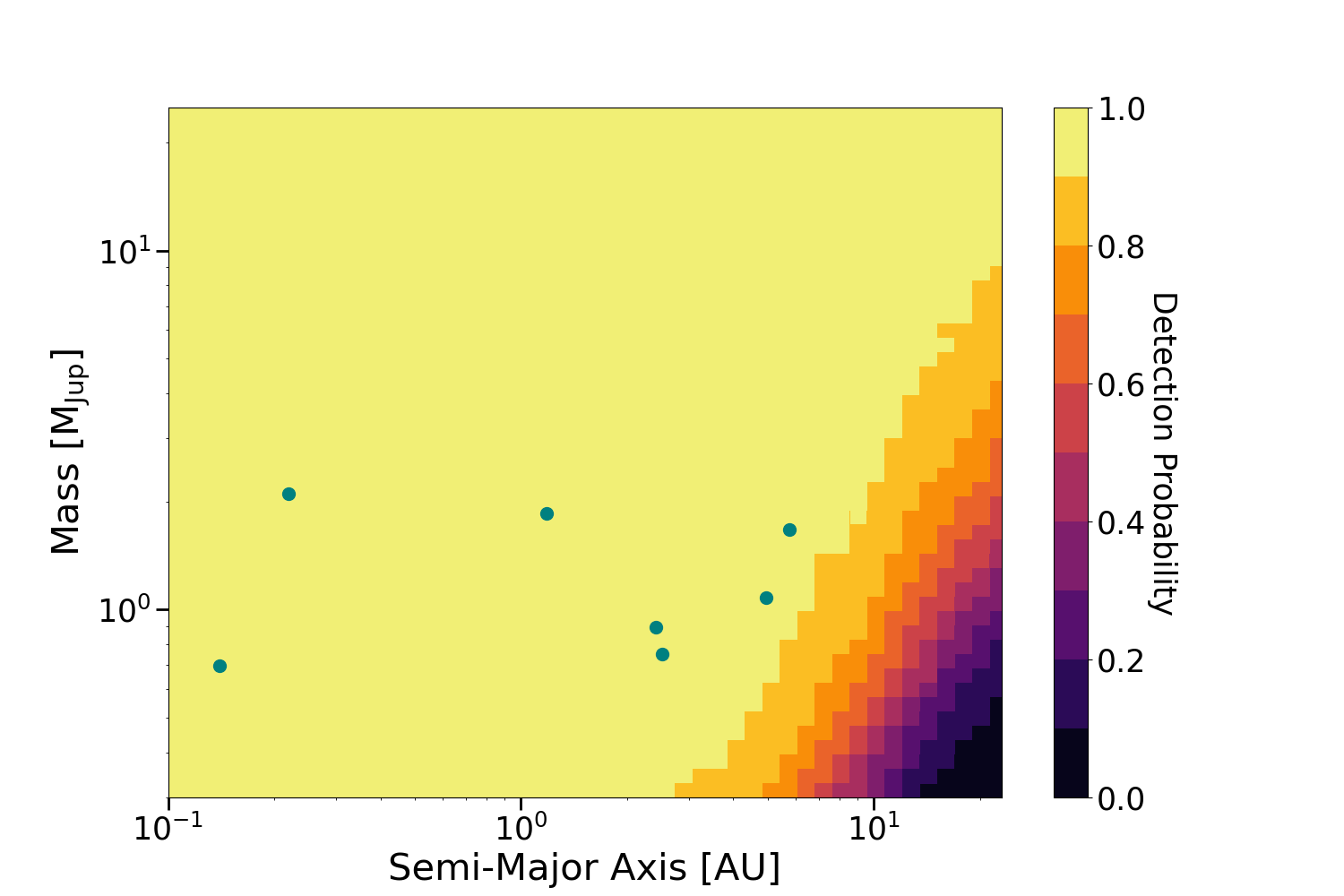} %&
    \includegraphics[width=0.5\textwidth]{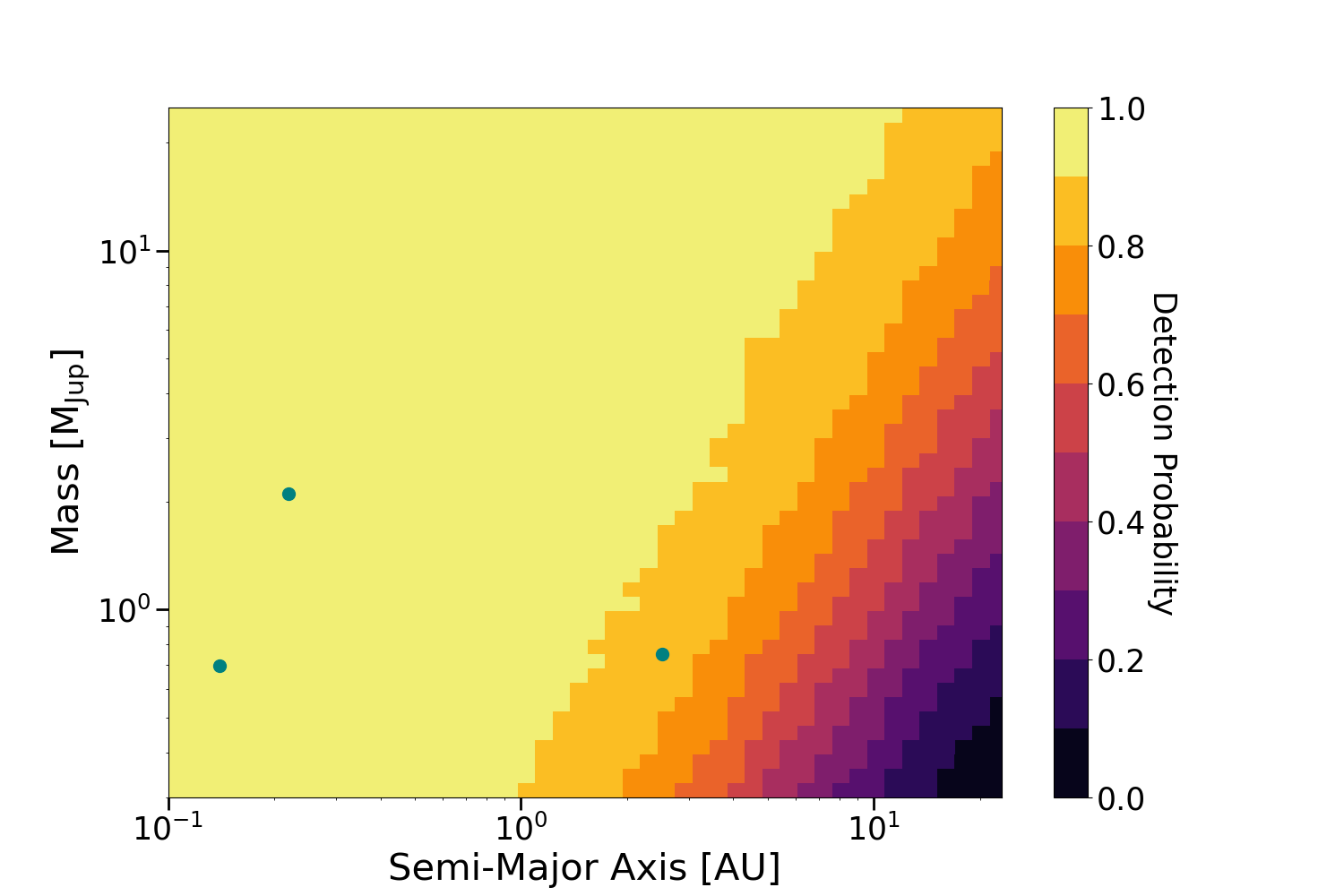}%\\
    %\end{tabular}
    \caption{Left: average completeness map for the R21 sample. Right: average completeness map for this paper's sample. Detected gas giants in each sample are overplotted in teal.}
    \label{fig:ave completeness}
\end{figure}

\bibliography{bibliography}{}

\end{document}